\documentclass[epj,twocolumn]{svjour}
\usepackage{latexsym}
\usepackage{graphics}
\usepackage{graphicx}
\usepackage{amsmath}
\usepackage{color}

\begin{document}

\title{Analytical study of level crossings in the Stark-Zeeman spectrum of ground state OH}
\author{N. Cawley \inst{1} \and Z. Howard \inst{1} \and M. Kleinert \inst{2} \and M. Bhattacharya \inst{1}
}                     
\institute{School of Physics and Astronomy, Rochester Institute of Technology,
84 Lomb Memorial Drive, Rochester, NY 14623, USA \and Department of Physics, Willamette University, 900 State Street, Salem,
OR 97301, USA}
\date{\today}
%
\abstract{
The ground electronic, vibrational and rotational state of the OH molecule is
currently of interest as it can be manipulated by electric and magnetic fields for experimental
studies in ultracold chemistry and quantum degeneracy. Based on our recent exact solution of the
corresponding effective Stark-Zeeman Hamiltonian, we present an analytical study of the crossings
and avoided crossings in the spectrum. These features are relevant to non-adiabatic transitions,
conical intersections and Berry phases. Specifically, for an avoided crossing employed in the
evaporative cooling of OH, we compare our exact results to those derived earlier from perturbation theory.
} 
\maketitle

\section{Introduction}
\label{sec:Intro}
The OH molecule is currently the subject of wide theoretical as well as experimental investigation in the context
of quantum computation \cite{Lev2006}, precision measurement \cite{Hudson2006,Kozlov2009}, ultracold collisions
\cite{Avdeenkov2003,Ticknor2005,Sawyer2008,Tscherbul2010} and quantum degenerate fluids \cite{Quemener2012}.
The ground state of the molecule displays both electric as well as magnetic dipole moments.
This polar paramagnetic character of OH makes intermolecular interactions physically interesting
\cite{Quemener2013}, and ensures that electric and magnetic fields can be extensively used to slow, guide, and trap
ultracold OH molecules \cite{Bochinski2004,Meerakker2005,Sawyer2007,Stuhl2012}.

The effect of electric and magnetic fields on the OH molecule can be described by an eight dimensional effective
Stark-Zeeman Hamiltonian when hyperfine structure, spin-orbit coupling and electric quadrupole effects are negligible,
such as at strong fields or high molecular temperatures \cite{Stuhl2012}. This Hamiltonian has been successfully
used to numerically model experimental data \cite{Quemener2012,Stuhl2012}. There have also been efforts towards
obtaining analytical solutions to the Hamiltonian, as exemplified by the exact diagonalization of its
field-dependent part \cite{Bohn2013}. Recently, our group presented the full analytical solutions
for the $X^{2}\Pi_{3/2}$ OH ground state Hamiltonian in combined electric and magnetic fields, neglecting
hyperfine structure. We also identified the underlying symmetry that enables the analytic solution
\cite{Mishkat2013}.

The most prominent features visible in the corresponding molecular spectrum are multiple
crossings and avoided crossings, which display rich behavior as the magnitudes and mutual orientation of the electric
and magnetic fields vary (see Fig.~\ref{fig:F1}), as has been shown in several experiments \cite{Quemener2012,Stuhl2012}.
These (avoided) crossings are
related to important physical phenomena such as Majorana transitions, responsible for OH trap loss
\cite{Lara2008}; Landau-Zener processes, which can be exploited for state transfer \cite{Stuhl2012}; evaporative
cooling, which is essential to Bose-Einstein condensation \cite{Quemener2012}; and conical intersections,
which provide pathways for molecular reactions \cite{Matsika2011}.

In the present work, we study analytically the crossings and avoided crossings in the spectrum of the OH ground
state Stark-Zeeman Hamiltonian, using algebraic techniques established previously
\cite{MishkatCrossingAJP,MishkatCrossingPRA1,MishkatCrossingPRA2}. We show that although the
(avoided) crossings display quite complex behavior, our analytical approach can organize and
characterize this behavior systematically. Focusing on the experimentally relevant situation where the
electric and magnetic field vectors are the tunable parameters \cite{Quemener2012,Stuhl2012}, we demonstrate
that the locations of a particular subset of the (avoided) crossings can be found analytically. We use this knowledge
to analyze in detail the gap at a specific avoided crossing important to the evaporative cooling of ground state
OH molecules \cite{Quemener2012} and compare our exact results to those derived earlier using perturbation theory
\cite{Stuhl2012}.

The remainder of this paper is organized as follows: In Section ~\ref{sec:Ham}, we present the Stark-Zeeman Hamiltonian.
We then discuss the crossings and avoided crossings in the spectrum in Section ~\ref{sec:Disc}. Lastly, we conclude in
Section ~\ref{sec:Conc}.

\section{Hamiltonian}
\label{sec:Ham}
The Hamiltonian of the $X^{2}\Pi_{3/2}$ OH molecule is given by \cite{Stuhl2012,Mishkat2013}
\begin{equation}
\label{eq:H1}
H=H_{o}-\vec{\mu}_{e}\cdot \vec{E}-\vec{\mu}_{b}\cdot \vec{B},
\end{equation}
where $H_{o}$ is the field-free Hamiltonian, $\vec{\mu}_{e}$ and $\vec{\mu}_{b}$ are the electric and magnetic
molecular dipole moments, respectively, and $\vec{E}$ and $\vec{B}$ are the electric and magnetic fields, respectively.
The matrix representation of this Hamiltonian has been obtained earlier in the literature as \cite{Stuhl2012}
\begin{equation}
\label{eq:Hmatrix}
H_M=
\left(
\begin{array}{cc}
A_{1}-A_{2} & -C\\
-C & A_{1}+A_{2}\\
\end{array} \right),
\end{equation}
where
\begin{align}
A_{1} =\frac{2}{5} \mu_B B
&\left( \begin{array}{cccc}
-3 & 0 & 0 & 0\\
0 & -1 & 0 & 0 \\
0 & 0 & 1 & 0 \\
0 & 0 & 0 & 3\\
\end{array} \right), \\
A_{2} = \frac{\hbar \Delta}{2} &\left( \begin{array}{cccc}
1 & 0 & 0 & 0\\
0 & 1 & 0 & 0 \\
0 & 0 & 1 & 0\\
0 & 0 & 0 & 1\\
\end{array} \right), \\
C = \frac{\mu_{e} E}{5} &\left( \begin{array}{cccc}
-3 \cos \theta &  \sqrt{3} \sin \theta & 0 & 0 \\
 \sqrt{3} \sin \theta & -\cos \theta & 2 \sin \theta & 0 \\
0 & 2 \sin \theta &  \cos \theta & \sqrt{3} \sin \theta \\
0 & 0 &  \sqrt{3} \sin \theta & 3 \cos \theta \\
\end{array} \right),\\
\nonumber
\end{align}
where $\Delta$ is the lambda-doubling parameter, $\mu_{B}$ is the Bohr magneton, $\mu_e$ is the magnitude
of the molecular electric dipole moment, $E=|\vec{E}|$ and $B=|\vec{B}|$ are the electric and magnetic field magnitudes,
respectively, and $\theta$ is the angle between the magnetic and electric field vectors. We note that
this Hamiltonian has been used successfully to describe several experiments \cite{Quemener2012,Stuhl2012}. We also note
that the validity of this Hamiltonian has two limitations: First, hyperfine as well as spin-orbit interactions have
been neglected, which is justified for ongoing experiments \cite{Stuhl2012}. Second, the effect of the electric quadrupole
term has been ignored. This term can become comparable to or larger than the magnetic dipole term (about 100 MHz at 100
Gauss) for a gradient in the electric field of a few percent.

\begin{figure*}[t]
\includegraphics[width=\textwidth]{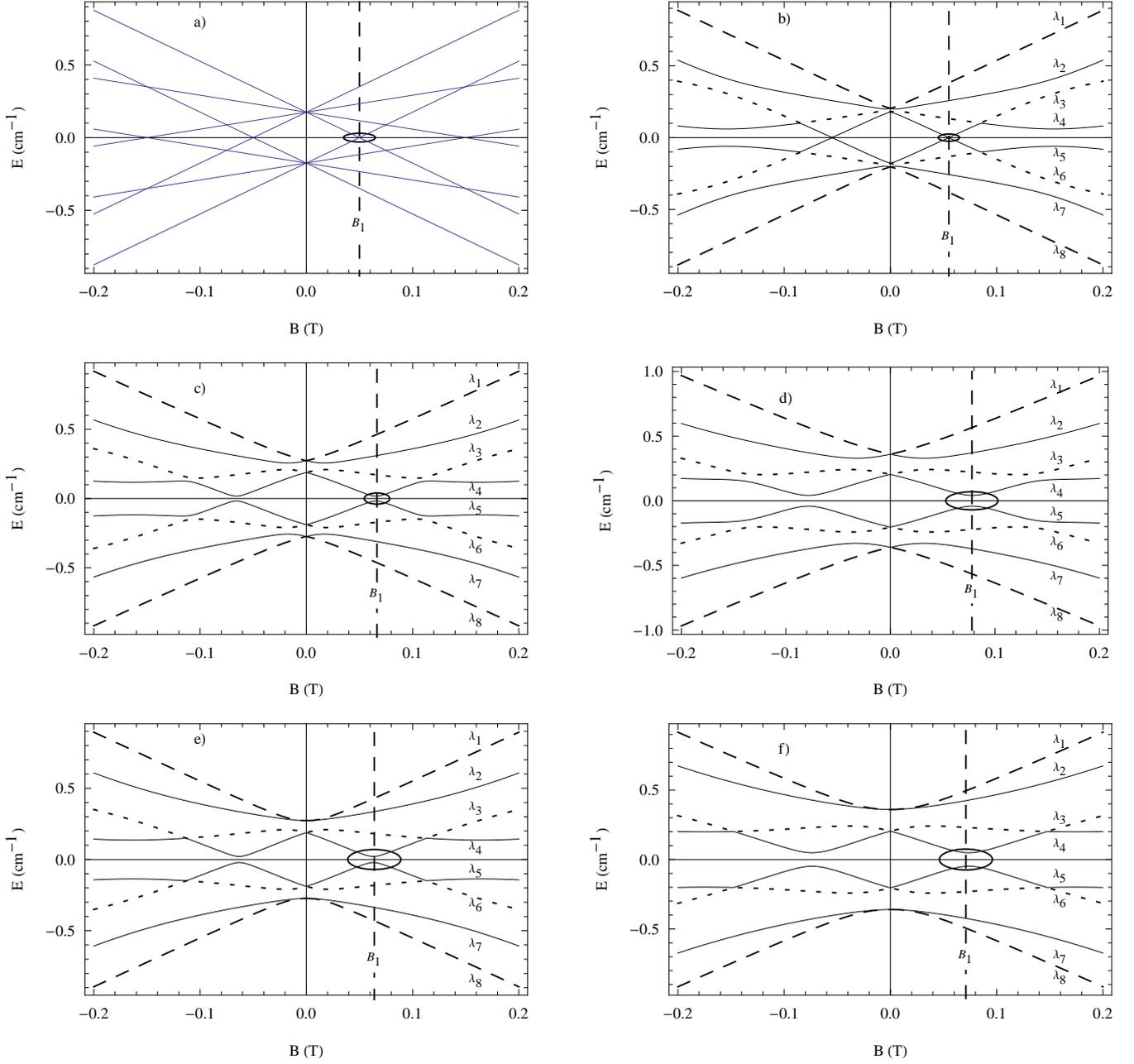}
\caption{Spectrum of the Hamiltonian of Eq.~(\ref{eq:Hmatrix}) for $\Delta=2\pi\times1.667$ GHz, $\mu_{e}=1.66$ D,
and a) $\theta=\pi/3$, $E=0$ kV/cm, b) $\theta=\pi/3$, $E=1$ kV/cm, c) $\theta=\pi/3$, $E=2$ kV/cm,
d) $\theta=\pi/3$, $E=3$ kV/cm, e) $\theta=\pi/2$, $E=2$ kV/cm, f) $\theta=\pi/2$, $E=3$ kV/cm. The horizontal
axis represents the magnetic field in Tesla while the vertical axis represents energy in cm$^{-1}$. The
analytic eigenvalues are labeled as $\lambda_{i}, i=1-8$, consistent with the ordering in the text. The magnetic field
location of the spectral feature of interest to this work, indicated by an ellipse, has been labeled as $B_{1}$.
In a) $B_{1}$ indicates the position of a crossing which in b) - f) turns into an avoided crossing.}
\label{fig:F1}
\end{figure*}

The exact eigenvalues of $H_{M}$ can be found readily using {\it Mathematica} and have been provided in our
earlier publication \cite{Mishkat2013}. In this article, the eigenvalues will be labeled as
$\lambda_{i}, i=1-8,$ as shown in Fig.~\ref{fig:F1} b) - f). We note that the analytical solutions
maintain their energy ordering for a fixed nonzero electric field $E$ and angle $\theta$, and varying magnetic field
strength.

\section{The discriminant}
\label{sec:Disc}
In principle, crossings in the spectrum, defined as the intersection between two eigenvalues, can be found by equating
two eigenvalues $\lambda_{i}$ and $\lambda_j$ for $i \neq j$, $i,j=1...8$. Likewise and similarly, avoided crossings,
defined as locations where the
separation between two eigenvalues goes through a local minimum, can be found. Generally, this kind of approach is
not very systematic, is tedious, and in view of the complexity of the eigenvalues and the multiplicity of tunable
parameters in the present problem \cite{Mishkat2013}, not particularly insightful.

We will instead employ the more elegant and powerful algebraic approach, elaborated earlier in a series of articles \cite{MishkatCrossingAJP,MishkatCrossingPRA1,MishkatCrossingPRA2},
which utilizes the discriminant of the characteristic polynomial of a matrix. The discriminant can always be obtained
analytically, is available as a standard function in most symbolic computation software packages, and contains rather
complete and accessible information about the crossings and avoided crossings in the spectrum as a function
of the parameters of the Hamiltonian.

Since an extensive exposition of the algebraic technique we use in this article is already available
in the literature, we will only summarize below the properties of the discriminant relevant to the present article.
\begin{enumerate}
\item The \textit{real roots} of the discriminant correspond to the locations of \textit{crossings} in the spectrum.
Therefore, the simultaneous intersection of $n$ eigenvalues leads to ${n \choose 2}=n(n-1)/2$ crossings.
\item The \textit{real parts of complex roots} of the discriminant correspond to the location of \textit{avoided crossings}
in the spectrum.
\item Every (avoided) crossing that occurs due to the tuning of a parameter $P$ of the Hamiltonian
\textit{contributes a factor quadratic in $P$} to the discriminant.
\end{enumerate}

The discriminant $D[H_{M}]$ of the matrix $H_{M}$ can be easily calculated either by using the
eigenvalues $\lambda_{i}$ of $H_{M}$
\begin{eqnarray}
\begin{array}{l}
\label{eq:Discdefn}
D[H_{M}]=
\displaystyle\prod_{i<j}^{8}(\lambda_{i}-\lambda_{j})^{2},\\
\end{array}
\end{eqnarray}
or the characteristic polynomial of $H_{M}$
\begin{equation}
\label{eq:CPoly}
P(\lambda)=|H_{M}-\lambda I|=\displaystyle\sum_{n=0}^{8}p_{n}\lambda^{n},
\end{equation}
whose coefficients $p_{n}$ have been published by us \cite{Mishkat2013}. The expressions for the discriminant are
lengthy and complicated, and are thus shown only in Appendix A. Nonetheless, some exact conclusions can be drawn from them.
We note, however, that analytic knowledge of the eigenvalues does not generally imply that the locations of all crossings
and avoided crossings can be found exactly.

\subsection{Factoring the discriminant}
\label{subsec:FactorDisc}
The discriminant $D[H_{M}]$ can be written as the product of three factors,
\begin{equation}
\label{eq:DiscFactor}
D[H_{M}]=f_{0}f_{1}f_{2}^2,
\end{equation}
whose explicit form has been provided in Appendix A.
We will consider these factors to be polynomials in $B$, in
order to find the avoided crossings as the magnetic field varies, following experiments \cite{Quemener2012,Stuhl2012}.
We note that we could just as easily consider them to be polynomials in the electric field $E$ or the angle $\theta$ to
find the avoided crossings as these two parameters are tuned.

The first factor of $D[H_{M}]$ is $f_{0}$, an eighth degree term in $B$. This term has eight real roots at $B=0$,
corresponding to four real crossings at $B=0$ as seen in the spectrum, Fig.~\ref{fig:F1} b) - f). The second term, $f_{1},$ is
an eighth-degree polynomial with only even order terms in $B$. It can thus be thought of as a quartic in $B^2$, making it
analytically solvable. We will discuss this term in more detail in Appendix B. The third factor, $f_{2}^{2}$, is the
square of a polynomial $f_{2}$ of degree 16, and also even in $B$. It does not seem to be generally solvable. We will
discuss $f_{2}$ in Section ~\ref{subsec:F2}.

The evenness (in $B$) can be traced to the presence of Kramer's degeneracy due to the time-reversal symmetry of the
system \cite{MishkatCrossingPRA2}. Reversing time implements the transformation
\begin{equation}
B \rightarrow -B,
\end{equation}
which leaves the spectrum invariant as can be seen in Fig.~\ref{fig:F1}. This symmetry causes the coefficients of the
characteristic polynomial Eq.~(\ref{eq:CPoly}) to be even in $B$ and hence also all terms in the discriminant. A
similar argument implies that the discriminant is also even in $E$. We note that in Appendix A we have used the variables
$\tilde{B}=4 \mu_{B} B, \tilde{E}=2 \mu_{e} E$ and $\tilde{\Delta} =5 \hbar \Delta$ to make the long expressions tidy.
In the remainder of the article we will use either $E, B, \Delta$ or $\tilde{E}, \tilde{B}, \tilde{\Delta}$, as appropriate.

\subsection{The magnetic field $B_{1}$}
\label{subsec:DetRoots}
The roots of $f_{1}$ can be found by treating it as a quartic in $B^2$, which is analytically solvable \cite{Merriman1892}.
From these solutions, we recover the eight roots of the original octic $f_{1}$ in the manner shown in Appendix B.
In general, these roots, and therefore the magnetic field crossings and avoided crossings they correspond to, show
rather complex behavior as a function of the parameters $E$ and $\theta$ as can be seen in Fig.~\ref{fig:F1}. For instance,
the positions of the (avoided) crossings depend on both parameters, and the angle determines whether or not a transition is
avoided.

\begin{figure}
\includegraphics[width=0.4\textwidth]{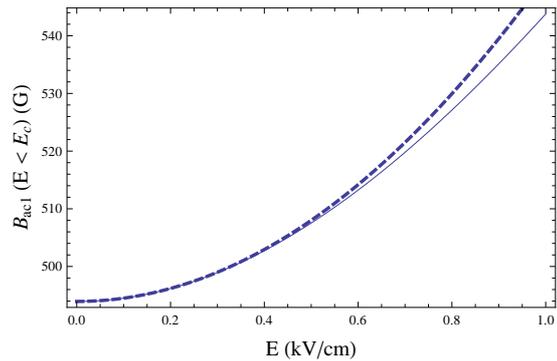}
\caption{Magnetic field location $B_{1}$ versus electric field $E$
for $\theta=\pi/3$. The solid line is the exact result of Eq.~(\ref{eq:Bless}), while the dotted line is the approximate expression
of Eq.~(\ref{eq:appBless}) for low electric fields.}
\label{fig:F2}
\end{figure}

\begin{figure}
\includegraphics[width=0.4\textwidth]{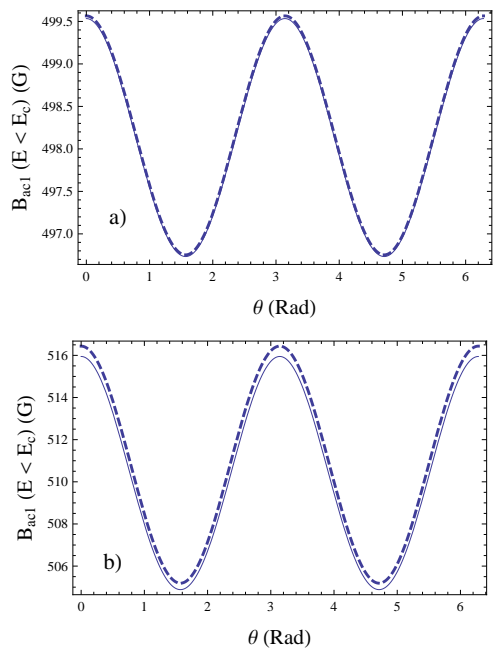}
\caption{Magnetic field location $B_{1}$ versus $\theta$ for a) $E=250$V/cm and b) $E=500$V/cm. The solid line is the exact
result of Eq.~(\ref{eq:Bless}), while the dotted line is the approximate expression of Eq.~(\ref{eq:appBless}) for low
electric fields.}
\label{fig:F3}
\end{figure}

In this article, we will focus only on one particular situation which is of experimental interest
\cite{Quemener2012,Stuhl2012} and concerns the circled crossing at $B_{1}$ in Fig.~\ref{fig:F1} a), where
\begin{equation}
B_{1}=\pm \frac{5\hbar\Delta}{12\mu_{B}},
\end{equation}
which is a real root of $f_{1}$. In Fig.~\ref{fig:F1} a) this crossing is seen to occur for $E=0$ at the magnetic field of
$\sim 0.05$T. In the presence of an electric field,  this crossing turns into an avoided crossing between the eigenvalues
$\lambda_{4}$ and $\lambda_{5}$, see Fig.~\ref{fig:F1} b) - f). The location $B_{1}$ of this avoided crossing has
been derived in Appendix B and depends strongly on the electric field strength and the angle between the electric and magnetic field
vectors. This dependence on the electric field strength was used experimentally to remove energetic
molecules from an OH trap to implement evaporative cooling \cite{Quemener2012}. We note that this magnetic field location
is also a function of $\Delta$, which is not tunable. In Fig.~\ref{fig:F2} and Fig.~\ref{fig:F3}, respectively, we compare
the variations of the exact analytical result for $B_{1}$ [Eq.~(\ref{eq:Bless})] with $E$ and $\theta$ to a simple
approximation valid for low electric fields
\begin{equation}
\tilde{B}_{1} \simeq \frac{\tilde{\Delta}}{3}+\frac{3(3 + \cos (2\theta)) \tilde{E}^2}{8 \tilde{\Delta}},
\end{equation}
that we derive in Appendix B [Eq.~(\ref{eq:appBless})]. As can be seen from Fig.~\ref{fig:F2} and Fig.~\ref{fig:F3}, this approximation
is quite good for electric fields less than $500$V/cm.

\subsection{The $\delta_{-3/2}$ gap}
The energy gap between $\lambda_{4}$ and $\lambda_{5}$ at $B_{1}$ was designated as $\delta_{-3/2}$ in
an experimental study of Landau-Zener losses in an OH trap \cite{Stuhl2012}, see Fig.~\ref{fig:F1}. Ground state OH molecules
can be removed controllably from their confining trap through the $\delta_{-3/2}$ gap by tuning the direction and magnitude
of an electric field. Thus, the scaling of the gap with $E$ and $\theta$ is of interest to current experiments
\cite{Quemener2012,Stuhl2012}.

In the present article, we find the size of the gap analytically by inserting the value of the magnetic field
$B_{1}$ into the eigenvalues $\lambda_{4}$ and $\lambda_{5}$,
\begin{equation}
\label{eq:gap}
\delta_{-3/2}=\lambda_{4}-\lambda_{5}=2\lambda_{4}(E,\theta,\Delta),
\end{equation}
where we have used the fact that $\lambda_{5}=-\lambda_{4}$ to express the gap in terms of a single eigenvalue,
$\lambda_{4}$. In the last step of Eq.~(\ref{eq:gap}), we have emphasized that $\delta_{-3/2}$ is only a function of
$E, \theta$ and $\Delta$ since the magnetic field $B_{1}$ depends on those parameters.

\begin{figure}
\includegraphics[width=0.4\textwidth]{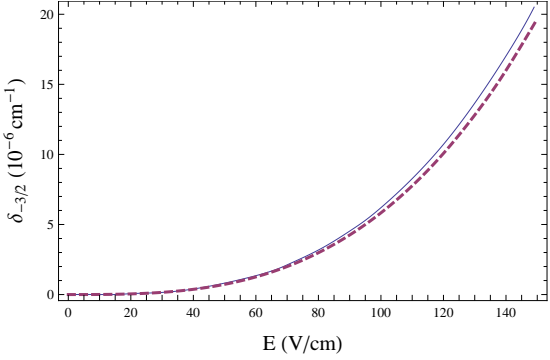}
\caption{Energy gap $\delta_{-3/2}$ versus electric field $E$ for $\theta = \pi/3$. The solid line
is the analytical result of Eq.~(\ref{eq:gap}), and the dotted line is a fit to $|a E^{3}|,$ (yielding $a =5830$) of the form in Eq.~(\ref{eq:dscaling}), as found earlier by perturbation theory \cite{Stuhl2012}.}
\label{fig:F5}
\end{figure}

\begin{figure}
\includegraphics[width=0.5\textwidth]{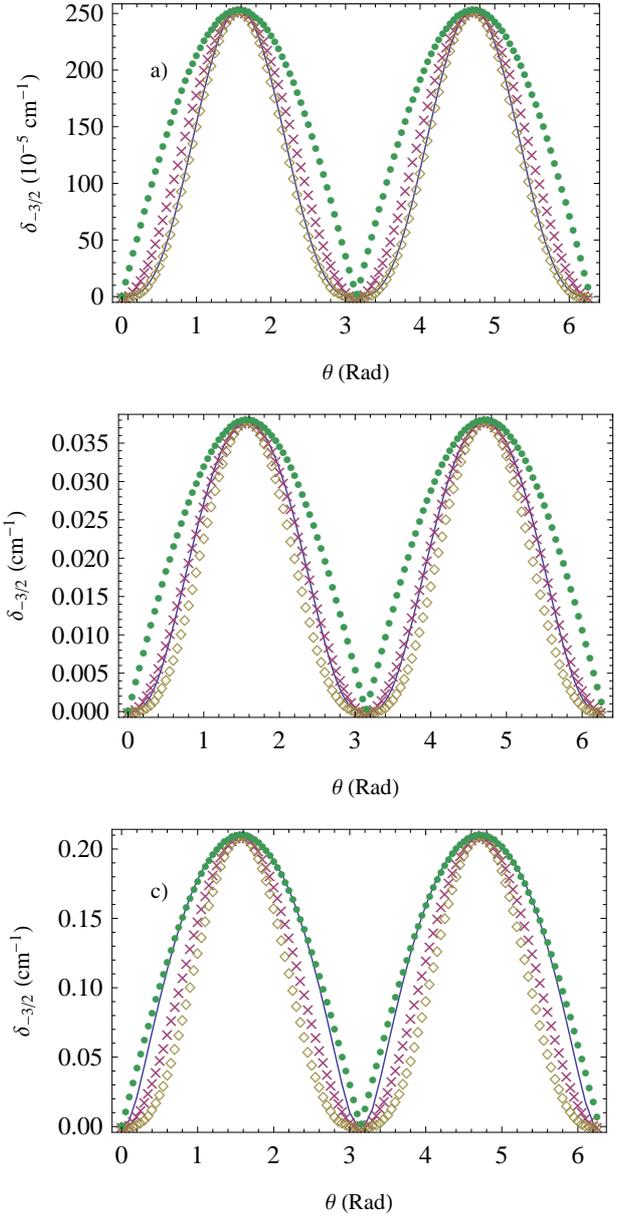}
\caption{Energy gap $\delta_{-3/2}$ versus $\theta$ for a) $E=300$V/cm, b) $E=1.4$kV/cm and c) $E=4$kV/cm. In the plots,
the solid line is the analytical result of Eq.~(\ref{eq:gap}), the solid dots correspond to a $|\sin\theta|$ curve,
the crosses correspond to a $\sin^{2}\theta$ curve and the diamonds correspond to a $|\sin^{3}\theta|$ curve of the form
in Eq.~(\ref{eq:dscaling}) as found earlier by perturbation theory \cite{Stuhl2012}. All curves are scaled to the
amplitude of analytical result. In a) the $|\sin^{3}\theta|$ curve matches well the exact result, in b) the
$\sin^{2}\theta$ is more appropriate, while in c) the $|\sin \theta|$ agrees best with the analytic answer.}
\label{fig:F6}
\end{figure}

In Fig.~\ref{fig:F5}
and Fig.~\ref{fig:F6},
we compare the field and angle scaling of our exact results to the result derived earlier using perturbation theory
\cite{Stuhl2012},
\begin{equation}
\label{eq:dscaling}
\delta_{-3/2} \propto |E^{3}\sin^{3}\theta|.
\end{equation}
From Fig.~\ref{fig:F5} we see that the perturbative scaling $\delta_{-3/2} \propto |E^{3}|$ is valid for
electric fields less than about $70$V/cm for $\theta=\pi/3$. From Fig.~\ref{fig:F6} a), we infer that the perturbative
scaling $\delta_{-3/2} \propto |\sin^{3}\theta|$ is accurate for low fields such as $300$V/cm. Our analytic results allow us
to explore higher electric fields, at which the gap $\delta_{-3/2}$ shows rather different scaling behavior with respect to
the angle $\theta$. In our calculations, for $E \geq 500$V/cm the variation of $\delta_{-3/2}$ with $|\sin\theta|$ deviates significantly
from the cubic law of Eq.~(\ref{eq:dscaling}). For instance, from Fig.~\ref{fig:F6} b) we see that at $E = 1.4$kV/cm, the
behavior is much better described by a quadratic dependence,
i.e. $\delta_{-3/2} \propto \sin^{2}\theta,$ while in Fig.~\ref{fig:F6} c) we see that at $E = 4$kV/cm, the analytic curve
follows closely the variation $\delta_{-3/2} \propto |\sin\theta|.$

\subsection{The determinant}
\label{subsec:DetCalc}
We note that $f_1$ can be written as
\begin{equation}
\label{eq:DiscF1}
f_{1}=N_a (\lambda_{1}-\lambda_{8})^{2}(\lambda_{2}-\lambda_{7})^{2}(\lambda_{3}-\lambda_{6})^{2}(\lambda_{4}-\lambda_{5})^{2},
\end{equation}
where $N_a=5^{8}$. This relation can be verified by using the analytical eigenvalue solutions $\lambda_{i}$
\cite{Mishkat2013}. Equation (\ref{eq:DiscF1}) implies that the zeros of $f_{1}$ correspond to (avoided)
crossings {\it only} between opposite energy pairs $\lambda_{i}$ and $\lambda_{9-i}$ in the spectrum. In general, these
are the only (avoided) crossings whose locations can be obtained analytically. Further, by using the relation \cite{Mishkat2013}
\begin{equation}
\label{eq:SymmC}
\lambda_{i} =-\lambda_{9-i},\,\,\, i=1,\ldots 8
\end{equation}
together with Eq.~(\ref{eq:DiscF1}), we find that
\begin{equation}
f_{1}=N_{b}\prod\limits_{j=1}^{8}\lambda_{j}=N_{b}|H_{M}|,
\end{equation}
where $N_{b}=10^{8}.$ Thus, $f_{1}$ is proportional to the determinant $|H_{M}|$ of the matrix $H_{M}$. Therefore, at every
crossing implied by the real roots of $f_{1}, |H_{M}|$ vanishes, which implies that at least two of the eigenvalues
have zero energy. As a check, in the case of the $\delta_{-3/2}$ gap considered above, it can be verified that at $E=0$
the two eigenvalues which cross, and which have zero energy, are $\lambda_{4}$ and $\lambda_{5}$, see Fig.~\ref{fig:F1} a).

\subsection{The roots of $f_{2}$}
\label{subsec:F2}
The last factor in the discriminant $D[H_{M}]$ is $f_{2}^{2}.$ All (avoided) crossings accounted for by $f_{2}$ occur
between levels which are \textit{not} opposite energy pairs. The polynomial is of order 16 in $B$. However,
since it is even in $B$, it can be thought of as an octic in $B^{2}$. This octic does not seem to be solvable
in general. This implies that the magnetic field locations such as for the avoided crossing gaps labeled $\delta_{\pm 1/2}$ in
earlier work \cite{Stuhl2012} cannot be found analytically. A detailed analysis of the solvability
of $f_{2}$ requires using the condition that the corresponding Galois group is solvable \cite{BasuBook}. However, in view of
the complexity of the coefficients, the result is not likely to be very illuminating, and we have not pursued this avenue.

The polynomial $f_{2}$ does yield analytic roots for some special cases. The simplest case, a trivial one, occurs when
$E=0.$ In that case,
\begin{equation}
f_2= 512 \tilde{B}^8 \tilde{\Delta}^4 \left(4 \tilde{B}^4 - 5 \tilde{B}^2 \tilde{\Delta}^2 + \tilde{\Delta}^4\right)^2, \nonumber
\end{equation}
corresponding to crossings at
\begin{equation}
B_{c}=0, \pm \frac{5\hbar\Delta}{8\mu_{B}}, \pm \frac{5\hbar\Delta}{4\mu_{B}}.\nonumber
\end{equation}
Less trivial, and possibly experimentally relevant solvable configurations occur when the electric and magnetic fields are
(anti)parallel [$\theta=(\pi)0$] or perpendicular $(\theta=\pi/2)$. In these cases, $f_2$ reduces to
\begin{align}
f_{2}[\theta=0]&=512 \left(\tilde{\Delta}^4 + 10 \tilde{\Delta}^2 \tilde{E}^2 + 9 \tilde{E}^4 \right) \left(4 \tilde{B}^8 + 4 \tilde{E}^8 \right. \nonumber\\
&\left. - 5 \tilde{B}^6 \left(\tilde{\Delta}^2 + 5 \tilde{E}^2\right) - 5 \tilde{B}^2 \tilde{E}^4 \left(\tilde{\Delta}^2 + 5 \tilde{E}^2\right) \right. \nonumber\\
&\left. + \tilde{B}^4 \left(\tilde{\Delta}^4 + 10 \tilde{\Delta}^2 \tilde{E}^2 + 42 \tilde{E}^4\right) \right)^2,\nonumber\\
f_{2}\left[\theta=\frac{\pi}{2}\right]&= 512 \tilde{B}^4 \tilde{\Delta}^4 \left(-4 \tilde{B}^2 + \tilde{\Delta}^2 + 8 \tilde{E}^2\right)^2 \left(\tilde{B}^8 + \tilde{E}^8 \right. \nonumber\\
&\left. - 2 \tilde{B}^6 \left(\tilde{\Delta}^2 - 2 \tilde{E}^2\right) + \tilde{B}^2 \tilde{E}^4 \left(\tilde{\Delta}^2 + 4 \tilde{E}^2 \right) \right. \nonumber\\
&\left. + \tilde{B}^4 \left(\tilde{\Delta}^4 + 8 \tilde{\Delta}^2 \tilde{E}^2 + 6 \tilde{E}^4\right)\right), \nonumber\\
f_{2}\left[\theta=\pi\right]&=512 \left(\tilde{\Delta}^4 + 10\tilde{\Delta}^2 \tilde{E}^2 + 9 \tilde{E}^4\right) \left(4 \tilde{B}^8 + 4 \tilde{E}^8 \right. \nonumber\\
&\left. - 5 \tilde{B}^6 \left(\tilde{\Delta}^2 + 5 \tilde{E}^2\right) - 5 \tilde{B}^2 \tilde{E}^4 \left(\tilde{\Delta}^2 + 5 \tilde{E}^2\right) \right. \nonumber\\
&\left. + \tilde{B}^4 \left(\tilde{\Delta}^4 + 10 \tilde{\Delta}^2 \tilde{E}^2 + 42 \tilde{E}^4\right)\right)^2,
\end{align}
for $\theta=0, \pi/2,$ and $ \pi$, respectively, and the locations of \textit{all} the crossings and avoided crossings in the spectrum of
$H_{M}$ can be found analytically in a straightforward way.

\section{Conclusion}
\label{sec:Conc}
We have presented an analytical level-crossing study of the $X^{2}\Pi_{3/2}$ state of the OH molecule, which is relevant to
ongoing experiments. We have clarified the set of crossings and avoided crossings whose locations can be found analytically.
We have analyzed a particular avoided crossing which is important to nonadiabatic transitions leading to
evaporative cooling of OH molecules towards Bose-Einstein condensation. We have derived exact as well as approximate
analytic expressions for the magnetic field location of this avoided crossing. We have also found an analytic expression
for the gap at the avoided crossing, and compared its scaling properties with respect to electric field and angle,
with results derived earlier from perturbation theory.

\appendix

\section{Discriminant}

The three factors in Eq.~(\ref{eq:DiscFactor}) are
\begin{align*}
f_{0}&=\left(\frac{\sqrt{3}}{156250 (2^{1/4})}\right)^8 \tilde{B}^8,\\
f_{1}&=81 \tilde{B}^8- 36 \left[ 9 \cos(2\theta) \tilde{E}^2 +5\tilde{\Delta}^2\right]\tilde{B}^{6}  +2 \left[59 \tilde{\Delta}^4 \right. \nonumber\\
&\left.+ 81 \left(2 + \cos(4\theta)\right) \tilde{E}^4 +54 \left(7-2\cos(2\theta)\right) \tilde{\Delta}^{2}\tilde{E}^{2}\right]\tilde{B}^{4}\nonumber\\
&-4 \left(\tilde{\Delta}^{2}+ 9 \tilde{E}^2 \right)\left[5 \tilde{\Delta}^{4}+\left(9 \tilde{E}^2-7 \tilde{\Delta}^2 \right) \tilde{E}^2 \cos(2 \theta) \right. \nonumber\\
&\left. +21 \tilde{\Delta}^{2}\tilde{E}^{2}\right]\tilde{B}^{2}+\left( \tilde{\Delta}^{4}+9 \tilde{E}^{4}+10 \tilde{\Delta}^2 \tilde{E}^2\right)^2,\\
f_{2}&=g_{16} \tilde{B}^{16} + g_{14} \tilde{B}^{14} + g_{12} \tilde{B}^{12} + g_{10} \tilde{B}^{10} + g_8 \tilde{B}^{8} \nonumber\\
&+ g_6 \tilde{B}^{6} +g_4 \tilde{B}^{4} + g_2 \tilde{B}^{2} +g_0,\\
\end{align*}

where $\tilde{\Delta}=5 \hbar\Delta, \tilde{B}=4 \mu_{B}B$ and $\tilde{E}=2 \mu_{e}E$ and

\begin{align*}
g_{16}&=8192 \left[\Delta^4 + 5 \left(1 + \cos(2 \theta)\right) \Delta^2 \tilde{E}^2 + 72 \cos^4 \theta \tilde{E}^4 \right], \nonumber\\
g_{14}&=-2048 \left[\right. 9 \cos^4 \theta \left(9 + 41 \cos(2 \theta) \right) \tilde{E}^6 + 10 \tilde{\Delta}^6  \nonumber\\
&\left. +\cos^2 \theta \left( 247 + 343 \cos(2 \theta) \right) \tilde{\Delta}^2 \tilde{E}^4 + 150 \cos^2 \theta \tilde{\Delta}^4 \tilde{E}^2 \right],\nonumber\\
g_{12}&=64 \left[\right. 264 \tilde{\Delta}^8 + 240 \left( 15 + 7 \cos(2 \theta) \right) \tilde{\Delta}^6 \tilde{E}^2\nonumber\\
&\left. + 2 \left( 7613 + 9308 \cos(2 \theta) + 1311 \cos(4 \theta) \right) \tilde{\Delta}^4 \tilde{E}^4 \right. \nonumber\\
&\left. + 9 \cos^4 \theta \left( 3155 + 2052 \cos(2 \theta) + 2481 \cos(4 \theta) \right) \tilde{E}^8 \right. \nonumber\\
&\left. + 4 \cos^2 \theta \left( 8599 + 13060 \cos(2 \theta) + 3501 \cos(4 \theta) \right) \tilde{\Delta}^2 \tilde{E}^6 \right],\nonumber\\
g_{10}&= -32 \left[160 \tilde{\Delta}^{10} +16 \left(203 + 47 \cos(2 \theta) \right) \tilde{\Delta}^8 \tilde{E}^2 + 4 \left(5685 \right. \right. \nonumber\\
&\left. \left. + 3884 \cos(2 \theta) + 631 \cos(4 \theta)\right) \tilde{\Delta}^6 \tilde{E}^4 + 36 \cos^4 \theta \left( 1620  \right. \right. \nonumber\\
&\left. \left. + 5367 \cos(2 \theta) +1188 \cos(4 \theta) + 1025 \cos(6 \theta) \right) \tilde{E}^{10} \right. \nonumber\\
&\left. + 4 \cos^2 \theta \left( 39498 + 56409 \cos(2 \theta) + 27750 \cos(4 \theta) \right. \right. \nonumber\\
&\left. \left. + 2903 \cos(6 \theta) \right) \tilde{\Delta}^2 \tilde{E}^8  + \left( 72962 + 100955 \cos(2 \theta) \right. \right. \nonumber\\
&\left. \left. + 33550 \cos(4 \theta) + 4533 \cos(6 \theta) \right) \tilde{\Delta}^4 \tilde{E}^6 \right],\nonumber\\
g_8&=8 \left[64 \tilde{\Delta}^{12} +192 \tilde{\Delta}^{10} \tilde{E}^2 \left(9 + \cos(2 \theta) \right) \right. \nonumber\\
&\left. +8 \tilde{\Delta}^8 \tilde{E}^4 \left(2193 + 1012 \cos(2 \theta) +339 \cos(4 \theta)\right) \right. \nonumber\\
&\left. +16 \tilde{\Delta}^6 \tilde{E}^6 \left(5651 + 6444 \cos(2 \theta) + 3093 \cos(4 \theta) \right. \right. \nonumber\\
&\left. \left. + 252 \cos(6 \theta)\right)  +72 \tilde{E}^{12} \cos^4 \theta \left(8253 + 6804 \cos(2 \theta) \right. \right. \nonumber\\
&\left. \left. + 7786 \cos(4 \theta) + 900 \cos(6 \theta)  + 625 \cos(8 \theta)\right)  \right. \nonumber\\
&\left.  +4 \tilde{\Delta}^2 \tilde{E}^{10} \cos^2 \theta \left(199593 + 305817 \cos(2 \theta) \right. \right. \nonumber\\
&\left. \left. + 135562 \cos(4 \theta) + 38183 \cos(6 \theta) + 1165 \cos(8 \theta)\right) \right. \nonumber\\
&\left. + \tilde{\Delta}^4 \tilde{E}^8 \left(305959 + 533164 \cos(2 \theta) + 289236 \cos(4 \theta) \right. \right. \nonumber\\
&\left. \left. + 55892 \cos(6 \theta) + 3077 \cos(8 \theta)\right) \right],\\
g_6&=\left[- \tilde{\Delta}^{10} \left( 64+ 2816 \cos(2 \theta) +2240 \cos(4 \theta)\right) \right. \nonumber\\
&\left. -16 \tilde{\Delta}^8 \tilde{E}^2 \left(354 + 4215 \cos(2 \theta) + 3326 \cos(4 \theta) \right. \right. \nonumber\\
&\left. \left. + 105 \cos(6 \theta)\right) -1152 \tilde{E}^{10} \cos^4 \theta \left(1620 + 5367 \cos(2 \theta) \right. \right. \nonumber\\
&\left. \left. +1188 \cos(4 \theta) + 1025 \cos(6 \theta)\right) -64 \tilde{\Delta}^2 \tilde{E}^8 \cos^2 \theta \left(67824 \right. \right. \nonumber\\
&\left. \left. + 129141 \cos(2 \theta) + 44446 \cos(4 \theta) + 12779 \cos(6 \theta)\right. \right. \nonumber\\
&\left. \left.  - 1070 \cos(8 \theta)\right) -4 \tilde{\Delta}^6 \tilde{E}^4 \left(38821 + 159112 \cos(2 \theta) \right. \right. \nonumber\\
&\left. \left. + 117620 \cos(4 \theta) + 11768 \cos(6 \theta) - 921 cos(8 \theta)\right) \right. \nonumber\\
&\left. +\tilde{\Delta}^4 \tilde{E}^6 \left(-1413318   -3053506 \cos(2 \theta) - 1941176 \cos(4 \theta) \right. \right. \nonumber\\
&\left. \left. - 392525 \cos(6 \theta) + 11646 \cos(8 \theta) + 4879 \cos(10 \theta)\right )\right]\tilde{E}^{4},\nonumber\\
g_4&=4\left[1575 \tilde{\Delta}^8 +\tilde{\Delta}^8 \left(1616 \cos(2 \theta) + 844 \cos(4 \theta)\right)\right. \nonumber\\
&\left. +144 \tilde{E}^8 \cos^4 \theta \left(3155 + 2052 \cos(2 \theta) + 2481 \cos(4 \theta)\right)\right. \nonumber\\
&\left. +8 \tilde{\Delta}^2 \tilde{E}^6 \cos^2 \theta \left(91042 + 69141 \cos(2 \theta) + 52350 \cos(4 \theta) \right. \right. \nonumber\\
&\left. \left. - 11253 \cos(6 \theta)\right) +432 \tilde{\Delta}^8 \cos(6 \theta)+ \tilde{\Delta}^4 \tilde{E}^4 \left(198181 \right. \right. \nonumber\\
&\left. \left.  + 249080 \cos(2 \theta) + 118740 \cos(4 \theta) + 38536 \cos(6 \theta) \right. \right. \nonumber\\
&\left. \left. - 21113 \cos(8 \theta)\right) +2 \tilde{\Delta}^6 \tilde{E}^2 \left(15185 + 1675 \cos(2 \theta)\right. \right. \nonumber\\
&\left. \left.  + 8580 \cos(4 \theta) + 3856 \cos(6 \theta) - 2133 \cos(8 \theta)\right) \right. \nonumber\\
&\left. -243 \tilde{\Delta}^8 \cos(8 \theta)\right]\tilde{E}^{8},\nonumber\\
\end{align*}
\begin{align*}
g_2&=512 \cos^2 \theta \left[\tilde{\Delta}^6 \left( 3- 64 \cos(2 \theta)\right) -36 \tilde{E}^6 \cos^2 \theta \left(9 \right. \right. \nonumber\\
&\left. \left. + 41 \cos(2 \theta)\right) +21 \tilde{\Delta}^6 \cos(4 \theta) +2 \tilde{\Delta}^4 \tilde{E}^2 \left(-3 \right. \right. \nonumber\\
&\left. \left. - 436 \cos(2 \theta) + 139 \cos(4 \theta)\right)+4 \tilde{\Delta}^2 \tilde{E}^4 \left(-118 \right. \right. \nonumber\\
&\left. \left. - 655 \cos(2 \theta) + 183 \cos(4 \theta)\right)\right]\tilde{E}^{12},\nonumber\\
g_0&=4096 \tilde{E}^{16} \left(\tilde{\Delta}^2 + 9 \tilde{E}^2\right) \cos^2 \theta \left(5 \tilde{\Delta}^2 + \tilde{E}^2 \right. \nonumber\\
&\left. + \left(-3 \tilde{\Delta}^2 + \tilde{E}^2\right) \cos(2 \theta)\right).\\
\end{align*}

\section{Roots of $f_{1}$}
The roots of $f_{1}$ can be found using standard methods. Hence we supply only an outline of the derivation below.
To find the roots of $f_{1}$ it is convenient to divide by the coefficient of the first term to obtain a new polynomial
\begin{equation}
\label{eq:EqF1}
F_{1}=\frac{f_{1}}{81}= \tilde{B}^8 + c_{6}\tilde{B}^6+c_{4}\tilde{B}^4 + c_{2}\tilde{B}^2 + c_{0},
\end{equation}
where
\begin{align*}
c_{6} &=-\frac{20}{9}\tilde{\Delta}^{2}-4 \tilde{E}^{2}\cos 2\theta,\nonumber\\
c_{4} &= \frac{118}{81}\tilde{\Delta}^{4} +\frac{4}{3}\left(7-2\cos(2 \theta)\right)\tilde{E}^{2}\tilde{\Delta}^{2} + 2 \tilde{E}^4 \left( 2 +\cos(4 \theta) \right),\nonumber\\
c_{2} &=-\frac{4 \left(\tilde{\Delta}^{2}+9 \tilde{E}^{2}\right)}{81} \left(5\tilde{\Delta}^{4} +9 \cos(2\theta) \tilde{E}^4 \right. \nonumber\\
&\left. -7 \left( \cos(2 \theta) - 3 \right) \tilde{\Delta}^2 \tilde{E}^2 \right),\nonumber\\
c_{0} &= \frac{\left( \tilde{\Delta}^{4}+9 \tilde{E}^{4}+10 \tilde{\Delta}^{2}\tilde{E}^{2}\right)^2}{81}.\\
\nonumber
\end{align*}
The solutions to Eq.~(\ref{eq:dquartic}) are succinctly analyzed in terms of the corresponding depressed quartic
\cite{StegunBook,WikiQuartic}
\begin{equation}
\label{eq:dquartic}
f(u) = u^4 + qu^2 + ru + s,
\end{equation}
where
\begin{eqnarray}
u &=& \tilde{B}^{2}+ \frac{c_{6}}{4},\nonumber\\
q &=& c_{4}-\frac{3 c_{6}^2}{8},\nonumber\\
r &=& \frac{1}{8} (8 c_{2}-4 c_{4} c_{6}+c_{6}^3),\nonumber\\
s &=& c_{0}-\frac{1}{256} c_{6} (64 c_{2}-16 c_{4} c_{6}+3 c_{6}^3).\\
\nonumber
\end{eqnarray}
Since we are looking for complex roots of $F_{1},$ which correspond to avoided crossings,
we assume a root for Eq.~(\ref{eq:dquartic}) of the form $u=x+iy$, where $x$ and $y$ are both real.
With this assumption, the depressed quartic returns a root with real $x$ and $y$ only if the
discriminant of the resolvent cubic \cite{WikiQuartic} $\Delta_{c}$ obeys the relation
\begin{equation}
\Delta_{c}<0,
\end{equation}
where
\begin{equation}
\label{eq:Deltac}
\Delta_{c}=-\left(\frac{2}{3}\right)^{2}\left(\frac{32}{3}\right)^{9}
\left(\tilde{\Delta}\tilde{E}^{2}\right)^{4}\left(\tilde{\Delta}^{2}+9\tilde{E}^{2}\right)^{3}
\sin^{8}\theta G_{c},
\end{equation}
with
\begin{eqnarray}
\label{eq:Gc}
G_{c}&=&32 \tilde{\Delta}^6 + 16\tilde{E}^2 \tilde{\Delta}^{4} (25  - 23 \cos (2\theta)) \nonumber\\
&&+576 \tilde{E}^4 \tilde{\Delta}^2\sin^{2}\theta (7-\cos (2\theta))\nonumber \\
&& +2592\tilde{E}^6 \sin^{2}\theta \cos^{4}\theta.\nonumber\\
\end{eqnarray}
From Eq.~(\ref{eq:Deltac}) it follows that $\Delta_{c} <0$ if $G_{c}>0,$ which in turn
is readily apparent from Eq.~(\ref{eq:Gc}) as $G_{c}$ is the sum of four terms which are all positive.

In this case, the complex roots of the depressed [Eq.~(\ref{eq:dquartic})], as well as the original [Eq.~(\ref{eq:EqF1})]
quartic can be found analytically. It turns out that as the electric field $\tilde{E}$ is tuned, the real parts of the
two pairs of complex conjugate roots of Eq.~(\ref{eq:dquartic}) become degenerate at
\begin{equation}
C_{r}=0,
\end{equation}
where
\begin{align}
\label{eq:Cr1}
&C_{r}= \frac{2^{1/3}}{24 D_{B}} \left(2 q^2 + 24 s \right) + \frac{2^{2/3}}{24} D_{B} - \frac{q}{6},\nonumber\\
&D_{B}= 3^{1/3} \left(  2/3 q^3 + 9 r^2 - 24 q s  \right. \nonumber\\
&\left. + \sqrt{12 q^3 r^2 + 81 r^4 - 48 q \left( q^3 + 9 r^2 \right) s + 384 q^2 s^2 - 768 s^3} \right)^{1/3} .\nonumber\\
\end{align}
This occurs at the (angle-dependent) critical field
\begin{equation}
\tilde{E}_{c}=\frac{\tilde{\Delta}}{\sqrt{1-2\cos \left( 2\theta\right)}}.
\end{equation}
Physically, this corresponds (for $\tilde{B}>0$) to the coincidence of two avoided crossings between
the eigenvalues $\lambda_{4}$ and $\lambda_{5}$. Since we are interested always in the avoided crossing $\delta_{-3/2}$
occurring at the lower magnetic field, we find the real and imaginary parts of the roots of
Eq.~(\ref{eq:dquartic}) in terms of the squared variable $\tilde{B}_{1}'^{2}$ for $\tilde{E}<\tilde{E}_{c}$
to be
\begin{equation}
\label{eq:ReImBac1s}
\Re[\tilde{B}'^{2}_{1}] = -\sqrt{C_r}-\frac{c_6}{4},\,\,\,\Im[\tilde{B}'^{2}_{1}] = \sqrt{-\frac{r}{4 \sqrt{C_r}}+C_r+\frac{q}{2}}.\\
\end{equation}
Above the critical electric field ($\tilde{E} > \tilde{E}_{c})$ the corresponding expressions are instead
\begin{equation}
\label{eq:ReImBac2s}
\Re[\tilde{B}'^{2}_{1}]= \sqrt{C_r}-\frac{c_6}{4},\,\,\,\,\Im[\tilde{B}'^{2}_{1}]= \sqrt{\frac{r}{4 \sqrt{C_r}}+C_r+\frac{q}{2}}.\\
\end{equation}
Note the difference in sign for the $\sqrt{C_{r}}$ terms between Eqs.~(\ref{eq:ReImBac1s}) and (\ref{eq:ReImBac2s}).

Finally, the magnetic field location $\tilde{B}_{1}$ of the avoided crossing $\delta_{-3/2}$
below the critical field $\tilde{E}_{c}$ is readily found \cite{MostowskiBook}, i.e.
\begin{align}
\label{eq:Bless}
\tilde{B}_{1} &= \sqrt{\frac{\Re[\tilde{B}'^{2}_{1}]+\sqrt{\Re[\tilde{B}'^{2}_{1}]^2 + \Im[\tilde{B}'^{2}_{1}]^2}}{2}}. \nonumber \\
\end{align}
An approximate expression correct to lowest order in the electric field $\tilde{E}$ is found to be
\begin{equation}
\label{eq:appBless}
\tilde{B}_{1}(\tilde{E}<\tilde{E}_{c}) \simeq \frac{\tilde{\Delta}}{3}+\frac{3(3 + \cos (2\theta)) \tilde{E}^2}{8 \tilde{\Delta}}.
\end{equation}
In the text, Fig.~\ref{fig:F2} and Fig.~\ref{fig:F3} compare the approximation of Eq.~(\ref{eq:appBless})
to the exact result of Eq.~(\ref{eq:Bless}), showing the initial quadratic dependence with the electric field and the
$\cos 2\theta$ angular dependence. An expression for the magnetic field loation of the avoided crossing above the critical
electric field can also be found similarly.


\end{document}